\documentclass[useAMS,usenatbib]{mn2e}

\usepackage{psfig,epsfig,supertabular,wrapfig,placeins}
\usepackage{graphicx,amsmath,amssymb,subfigure,multirow,units}
\usepackage{marvosym}
\usepackage{graphics}
\usepackage{natbib}
\usepackage{amssymb}
\usepackage{amsmath}
\usepackage{euscript}
\usepackage{setspace}
\usepackage{fancyhdr}
\usepackage{afterpage,rotating,url}
\usepackage{times}

\newcommand{\Msolar}{\mbox{\,$\rm M_{\odot}$}}        
 
\newcommand{\asec}{\ensuremath{^{\prime\prime}}}

\newcommand{\object}{4C~+72.26}

\newcommand{\Hbeta}{H$\beta$}

\title[When galaxies collide: \object]{When galaxies collide:
  understanding the broad absorption-line radio galaxy \object}

\author[D.J.B.~Smith et al.]{D.J.B. Smith$^{1,2}$\thanks{E-mail: daniel.j.b.smith@gmail.com~(DJBS)}, C.~Simpson$^1$, A.M.~Swinbank$^3$, S.~Rawlings$^4$ \& M.J.~Jarvis$^5$ \\
    $^{1}$Astrophysics Research Institute, Liverpool John Moores University, Twelve Quays House, Egerton Wharf, Birkenhead, CH41 1LD, UK\\ 
    $^{2}$School of Physics and Astronomy, University of Nottingham, University Park, Nottingham, NG7 2RD, UK\\
    $^{3}$Institute for Computational Cosmology, Department of
  Physics, University of Durham, South Road, Durham DH1 3LE, UK \\
    $^{4}$Department of Astrophysics, University of Oxford, Denys Wilkinson Building, Keble Road, Oxford, OX1 3RH, UK \\
    $^{5}$Centre for Astrophysics, Science \& Technology Research Institute, University of Hertfordshire, Hatfield, Herts, AL10 9AB, UK }
    
\begin{document}

\date{\today}

\pagerange{\pageref{firstpage}--\pageref{lastpage}} \pubyear{2002}

\maketitle

\label{firstpage}

\begin{abstract}
We present a range of new observations of the `broad absorption line
radio galaxy' \object\ ($z \approx 3.5$), including sensitive
rest-frame ultraviolet integral field spectroscopy using the
Gemini/GMOS-N instrument and Subaru/CISCO $K$-band imaging and
spectroscopy. We show that \object\ is a system of two vigorously
star-forming galaxies superimposed along the line of sight separated
by $\sim$1300$\pm$200 km~s$^{-1}$ in velocity, with each demonstrating
spectroscopically resolved absorption lines. The most active
star-forming galaxy also hosts the accreting supermassive black hole
which powers the extended radio source. We conclude that the star
formation is unlikely to have been induced by a shock caused by the
passage of the radio jet, and instead propose that a collision is a
more probable trigger for the star formation. Despite the massive
starburst, the UV-mid-infrared spectral energy distribution suggests
that the pre-existing stellar population comprises $\sim
10^{12}$\Msolar\ of stellar mass, with the current burst only
contributing a further $\sim$2\%, suggesting that \object\ has already
assembled most of its final stellar mass.
\end{abstract}


\section{Introduction}

Powerful radio galaxies represent an important phase in the life of
all massive galaxies. The tight locus they follow in the $K$-band
Hubble diagram (Lilly \& Longair 1982; Jarvis et al.\ 2001; Willott et
al.\ 2003; Rocca-Volmerange et al.\ 2004) suggests that they are
luminous ($\sim3L^*$) galaxies whose stellar populations formed
rapidly at very high redshift and have evolved passively since. Their
descendants must inevitably be the most massive galaxies (with the
most massive black holes) and, locally, such galaxies lie in the
centres of galaxy clusters. During their most active formation phase,
they are believed to regulate galaxy formation and growth by heating
the intracluster medium with energy released from the continued
accretion of material onto their central supermassive black holes
(e.g. Best et al.\ 2006, Nesvadba et al., 2006, Nesvadba,
2009). However, most of the black hole growth must occur during a more
luminous episode (or episodes) of activity when the black hole powers
two oppositely-directed supersonic jets and has the characteristic
Fanaroff \& Riley (1974) Class~II double-lobed morphology. Although
the jets may be weak during most of the mass build-up
(e.g. Martinez-Sans\'igre et al., 2005), powerful episodes of jet
emission may cast their influence beyond the AGN's immediate
environment (e.g. Rawlings \& Jarvis, 2004). Understanding the
relationship between the star-formation and accretion activity is
essential to our understanding of massive galaxy evolution and
requires observations of distant radio sources.

High-redshift radio galaxies (HzRGs) have generally been the most
useful class of source, since their favourable orientation close to
the plane of the sky means the host galaxy's stellar population is not
swamped by the strong non-stellar continuum at ultraviolet and optical
wavelengths, and ensures the overall spectral energy distribution is
not affected by Doppler boosting of the synchrotron
emission. Archibald et al.\ (2001, and also Reuland et al. 2003) found
that the submillimetre luminosities of a sample of HzRGs were a
strongly increasing function of redshift, which they claimed was due
to increasingly youthful stellar populations (although this was
subsequently cast in to doubt by Rawlings et al., 2004). The apparent
lag between the assumed formation epoch ($z_{\rm f}=5$) and the
powerful AGN activity was explained by the time required for the
central black hole to grow to $10^9\rm\,M_\odot$ (Archibald et
al.\ 2002).

Deep spectroscopy of the $z=3.8$ radio galaxy 4C~41.17 by Dey et
al.\ (1997) revealed pronounced P~Cygni-like features in the
rest-frame ultraviolet which, coupled with the lack of polarization,
implies that the UV light is dominated by a massive starburst with an
age less than 16\,Myr. Considering the close spatial correspondence
between the radio, UV continuum, and Ly$\alpha$ emission, Bicknell et
al.\ (2000) suggested that 4C~41.17 was undergoing a radio jet-induced
starburst. While this scenario had been suggested before (Chambers et
al.\ 1987; McCarthy et al.\ 1987; Rees 1989), 4C~41.17 provided the
clearest observational example of the phenomenon. However, this
picture is clearly at odds with that proposed by Archibald et
al.\ (2002).

In this paper we discuss our observations of the $z\approx3.5$ source
\object, also known as 6C~1909+72, or TX~J1908+7220. Although
originially classified as a `broad absorption line radio galaxy' from
its optical spectrum spectrum (Dey 1999; de Breuck et al.\ 2001), we
demonstrate that the absorption features are P~Cygni profiles
characteristic of a massive starburst which provides the entire
rest-frame ultraviolet continuum.

The format of this paper is as follows. In Section~2 we describe the
new observations of \object, and in Section~3 we present and discuss
our major results, including the evidence for two star-forming systems
superimposed along the line of sight. In Section~4 we present a new
analysis of the spectral energy distribution of \object, and in
Section~5 we discuss our interpretation for the source. Finally, we
summarize our conclusions in Section~6. Throughout this paper we adopt
a cosmology with $H_0 = 72\rm\,km\,s^{-1}\,Mpc^{-1}$, and $\Omega_{\rm
  m}=1-\Omega_\Lambda=0.26$ (Dunkley et al.\ 2009).

\section{Observations}

\subsection{Imaging}

\begin{figure}
\centering
\includegraphics[width=0.90\columnwidth]{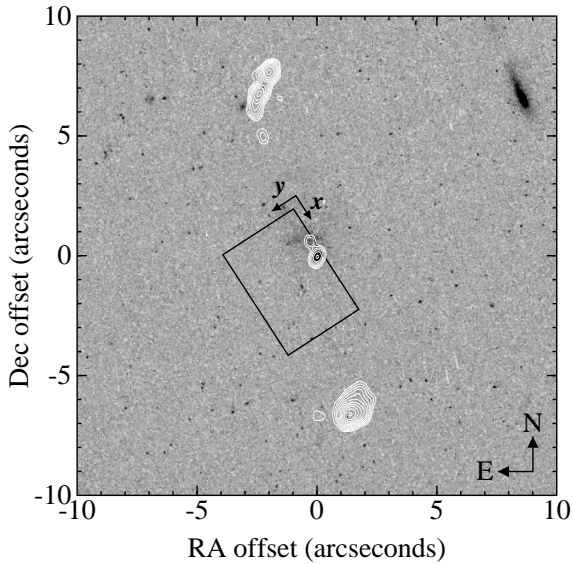}
\caption{{\it Hubble Space Telescope} WFPC2 Planetary Camera image of
  \object\ in the F702W filter, overlaid with the region covered by
  the GMOS-N IFU observations. The reason for the misplacement of the
  IFU is given in the text. Some residual cosmic rays are still
  visible in the image. The white contours are from the 4.5~G$Hz$
  radio map of Pentericci et al. (2000), kindly provided by C.~De
  Breuck. The radio core has been aligned with the peak of the optical
  emission. }
\label{fig:fov}
\end{figure}

Images of \object\ were taken through the $K'$ filter with the Cooled
Infrared Camera and Spectrograph for OHS (CISCO; Motohara et
al.\ 2002) on Subaru Telescope on the night of UT 2000 July 13. A
total of 64 separate 20-second exposures were taken, with 8 cycles of
an 8-point dither pattern. Due to CISCO's reset anomaly, the first
exposure in each cycle was discarded, giving a total exposure time of
1120\,s. These images were dark-subtracted, normalized, and
median-filtered to provide a flatfield which was applied to each
individual exposure.  These were then coadded using offsets calculated
from the centroids of objects detected in each frame. There was thin
cloud at the time of the observations so photometric calibration was
performed relative to three stars in the field which are present in
the Two Micron All Sky Survey (2MASS) catalogue. No colour corrections
were applied to transform from the CISCO $K'$ filter to the 2MASS
$K_{\rm s}$ filter since they have very similar transmission
profiles. The 5-$\sigma$ magnitude limit of these data is $K_s =
20.8$, measured in a 2\asec\ aperture.

We have also obtained an archival HST image of \object, taken using
the Wide-Field and Planetary Camera 2 with the F702W filter for two
3000-second integrations (PI G.~Miley). The two frames were combined
with cosmic-ray rejection using the \textsc{iraf/stsdas} task
\texttt{crrej} and the absolute flux scale was determined from
keywords in the image header. These data are presented in figure
\ref{fig:fov}.

\subsection{Longslit spectroscopy}
\label{subsec:kspec}

A $K$-band spectrum of \object\ was taken with CISCO on UT 2000 July
14. A 1-arcsec slit was used at a position angle of 33$^\circ$ to
align the slit with the extended radio structure (Pentericci et
al.\ 2000).  This setup provided a resolution
$\lambda/\Delta\lambda\approx300$ at a dispersion of
$\sim8.6$\,\AA\,pixel$^{-1}$. The source was acquired by blind
offsetting from a nearby USNO star; the offsets of 38.6\asec\ East and
6.4\asec\ North had been measured from the $K$-band image taken the
previous night. Twenty exposures of 200\,s each were taken, with the
target alternating between two positions along the slit separated by
10\,arcsec. One image in each pair was subtracted from the other to
provide first-order sky removal. Pixel-to-pixel sensitivity variations
were removed by dividing by the normalized $K'$ flatfield used in the
reduction of the imaging data. Curvature in the sky lines was
corrected by cross-correlating each row (spatial element) of the
longslit spectrum against the central row and shifting it by the
resulting amount. Residual sky emission was then removed by fitting a
second-order polynomial at each wavelength position. Wavelength
calibration was performed by fitting a second-order polynomial to
identified sky lines, with an rms of 16\,\AA.

A spectrum of the nearby F8 star SAO~9358 was taken and modelled as a
6140\,K blackbody to determine the relative sensitivty as a function
of wavelength, and the absolute flux scale derived from a
spectroscopic observation of the UKIRT Faint Standard FS~147 (Hawarden
et al.\ 2001). From cuts in the spatial direction of these spectra,
the seeing at the time of the observations was estimated to be
0.5\,arcsec.

\subsection{GMOS-N Integral Field Spectroscopy}

\object\ was observed using the GMOS-N instrument in IFU mode, between
30 May and 27 August 2006 under program ID GN-2006A-Q-36. The
instrument was used in single slit mode, providing a field of view of
$5''\times3.5''$ at a position angle 33$^\circ$ East of North, with
the R400 grating set to a central wavelength of 8037\,\AA. Again,
blind offsetting was required, and the Phase~2 document for the
observations provided the accurate relative offset from the USNO star
to the target, but included only nominal catalogue positions for the
offset star and target. However, the blind offsetting was performed by
resetting the telescope pointing after peaking up on the offset star,
and then moving to the \textit{absolute\/} astrometric coordinates of
the target. Since the USNO and 6C catalogues have different absolute
astrometric calibrations, this introduced a positional offset error of
1.6\asec\ which resulted in the core of the target lying close to the
edge of the instrument's field of view (Fig.~\ref{fig:fov}).

Ten 3600-second exposures were taken, and the data were reduced using
standard {\sc iraf} packages as well as making use of custom {\tt IDL}
routines specifically designed for the GMOS-N IFU (Swinbank et
al.\ 2003, 2006). The data were de-biased, flat-fielded, corrected for
the pixel-to-pixel CCD response, and wavelength calibrated based on
arc lamp exposures taken with the same instrumental setup at the same
time as the science observations. The sky lines were subtracted making
use of the sky fibers interspersed with the science fibers on the
combined CCD frames, and cosmic rays were rejected from the individual
frames, with additional flattening of each exposure making use of
empty regions of spectra to remove any remaining gradients in the
frames. The individual exposures were then mapped into a separate data
cube for each 3600s integration. The cubes were then combined to the
nearest ``spaxel'' making use of a 3$\sigma$ clipping algorithm, with
offsets based on a 2D fit to the continuum emission profile in each
data cube. The data were flux calibrated based on observations of the
spectrophotometric standard star Wolf~1346 using the same setup as for
the science observations, and the generated flux solution was applied
to the stacked cube. 

\section{Results}
\label{sec:Results}

\begin{figure}
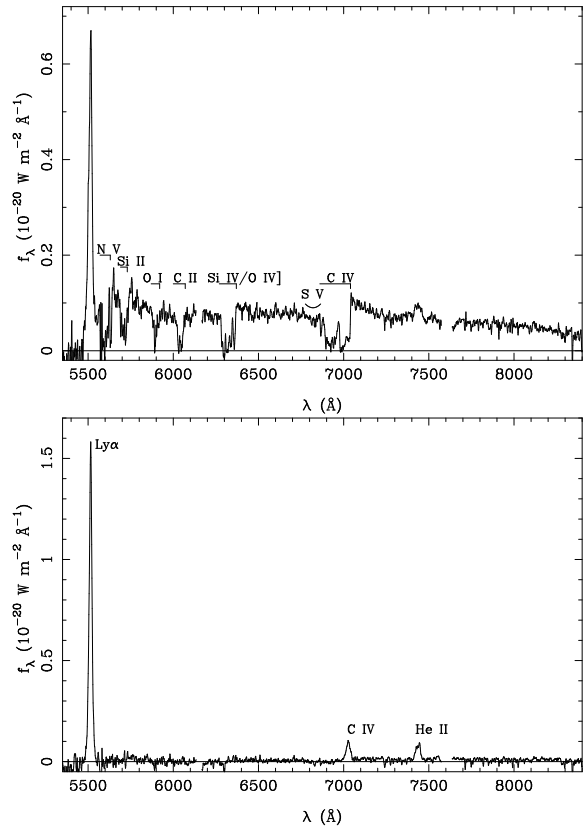

\centering
\rotatebox{-90}{\includegraphics[height=0.90\columnwidth]{core_spectrum.eps}}
\rotatebox{-90}{\includegraphics[height=0.90\columnwidth]{halo_spectrum.eps}}
\caption{GMOS-IFU spectra of \object\ extracted from $3\times3$-pixel
  regions around the continuum peak (top) and Ly$\alpha$ emission peak
  (bottom). Both spectra have been smoothed with a 3-pixel boxcar
  filter. The locations of emission lines are shown in the bottom
  panel, while the absorption features are labelled in the top
  panel. The two small gaps in the spectral coverage centred on
  $\sim$6150 and $\sim$7600\AA\ are due to the chip gaps on the GMOS
  CCDs.}
\label{fig:spectrum_both}
\end{figure}

\begin{figure}
\centering
\includegraphics[width=0.90\columnwidth]{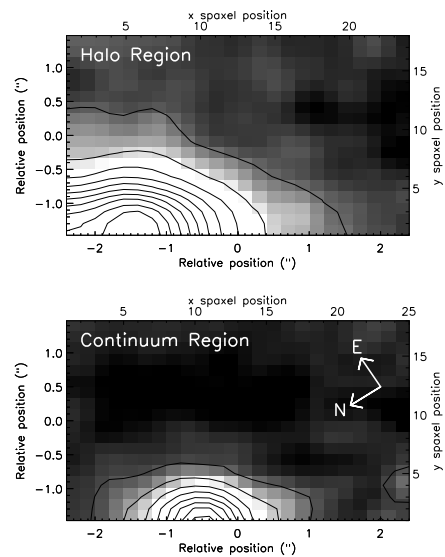}
\caption{Lyman-$\alpha$ narrow-band (top) and continuum (bottom)
  images from the data cube. It is clear that the extensive
  Lyman-$\alpha$ halo is offset from the peak intensity of the
  continuum emission. The observations were taken at a position angle
  of 33$^\circ$ East of North using the GMOS-N instrument in
  single-slit IFU mode. }
\label{fig:lyman_cont}
\end{figure}

We present optical spectra of \object\ in
Figs.~\ref{fig:spectrum_both}; whilst the upper spectrum appears very
similar to the long slit spectrum presented in De Breuck et al (2001),
our IFU data also have wavelength coverage up to $\sim 8400$\AA. In
addition to the narrow emission lines of Ly$\alpha$ and He{\sc~ii},
several deep broad absorption troughs can be seen. These are the
features which led Dey (1999) to refer to \object\ as a `broad
absorption line radio galaxy'.

In Fig.~\ref{fig:lyman_cont} we show two narrow-band images from our
stacked datacube. In the upper panel the Ly$\alpha$ emission is shown,
whilst the lower panel shows the spatial distribution of the continuum
between 7067\,\AA\ and 7211\,\AA\ (this region was chosen since it is
free of both absorption and emission line features). Although the
target is not centred in our field of view, we can see that the
Ly$\alpha$ emission has a greater extent than the continuum
emission. The different properties of these two regions are shown more
clearly in Fig.~\ref{fig:spectrum_both}, which shows spectra extracted
around the continuum peak (pixel 10,1)\footnote{We denote the
  bottom-left pixel of Fig.~\ref{fig:lyman_cont} as (0,0).} and the
Ly$\alpha$ emission peak (pixel 4,1). Comparing to
Fig.,~\ref{fig:fov}, we infer that the compact southern knot (which is
unresolved in the Planetary Camera image, implying a size $<500$\,pc)
is the site of most of the continuum emission, while the diffuse
emission seen to the north and northeast has a significant
contribution from the C{\sc~iv} and He{\sc~ii} emission lines. At the
very edge of our datacube, these lines are still visible but we do not
detect any continuum emission.

\subsection{The emission lines}
\label{subsec:emlines}

\begin{figure*}
\centering
\subfigure{\includegraphics[width=0.694\columnwidth]{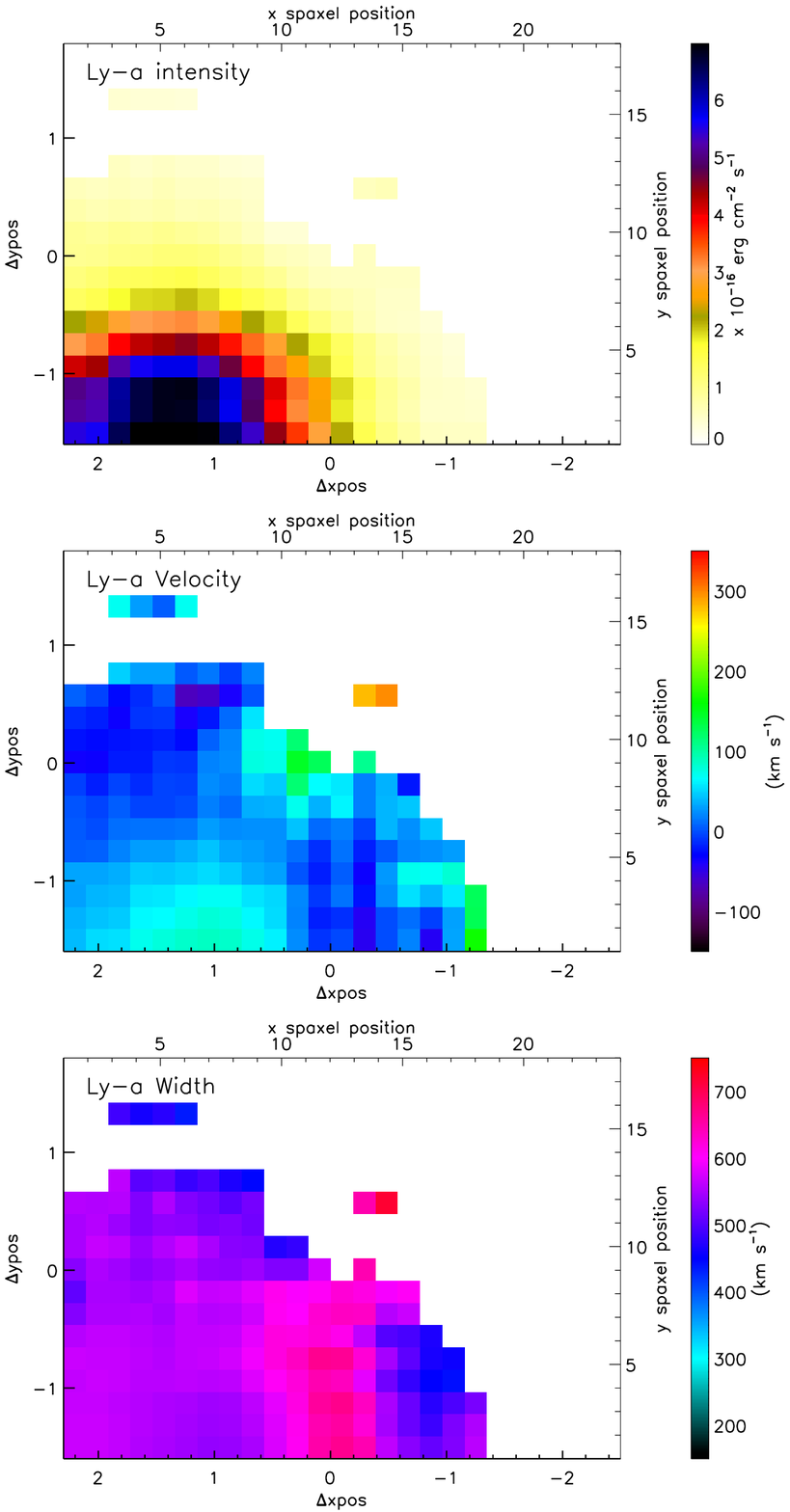}}
\subfigure{\includegraphics[width=0.694\columnwidth]{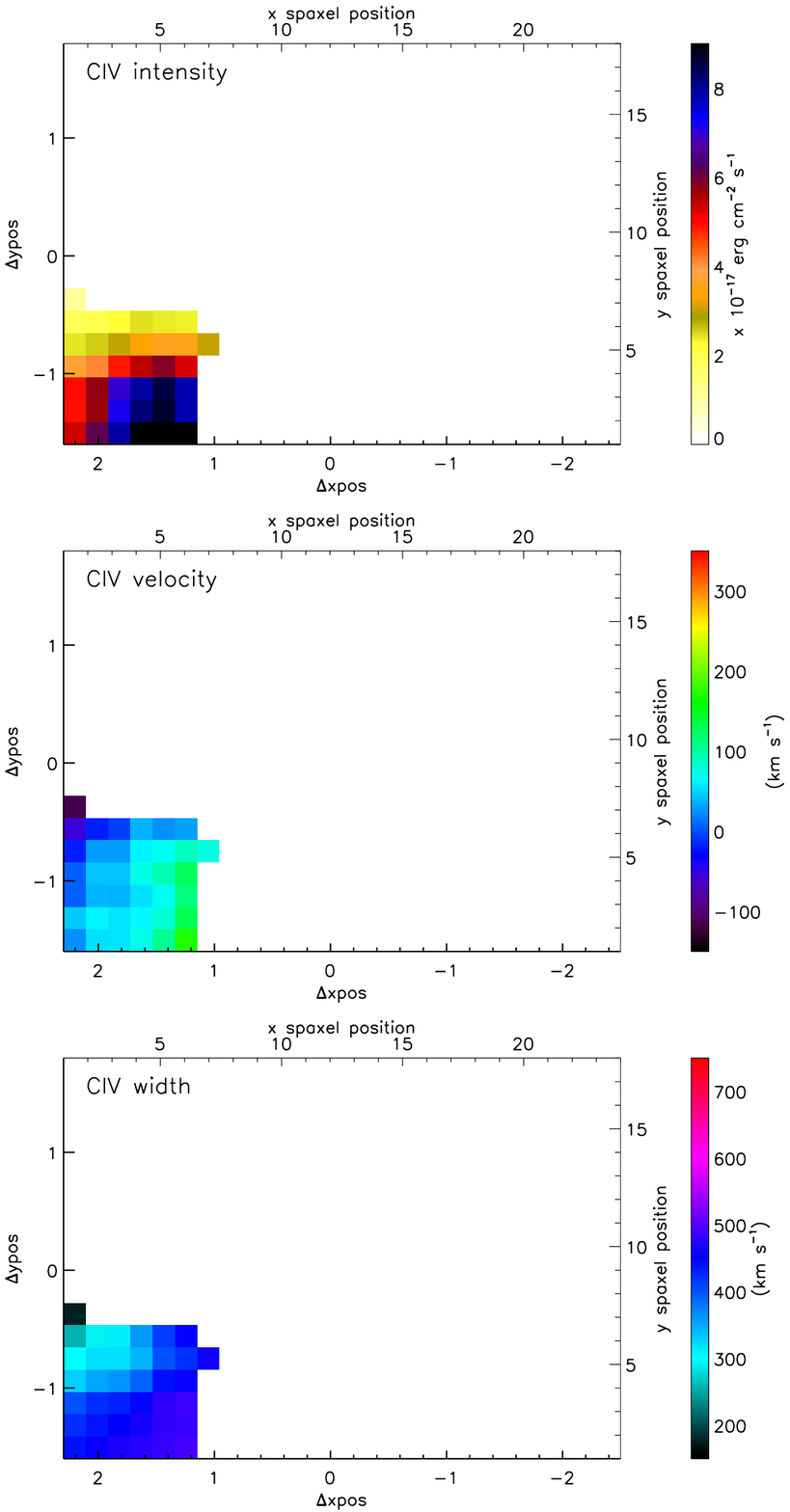}}
\subfigure{\includegraphics[width=0.694\columnwidth]{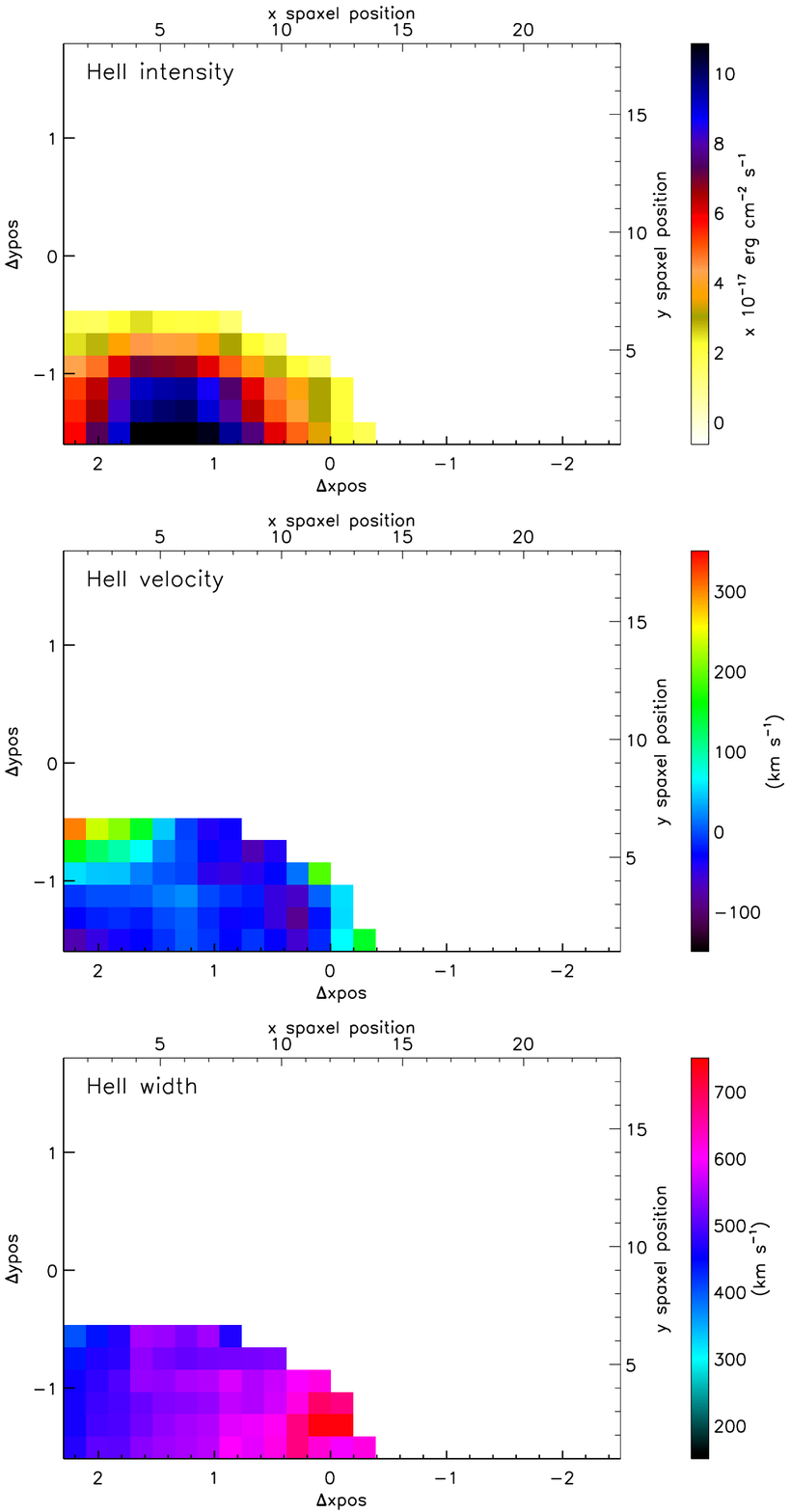}}
\caption{Lyman-$\alpha_{1216}$, and C{\sc~iv}$_{1549}$,
  He{\sc~ii}$_{1640}$ emission line intensity, velocity and linewidth
  as a function of position within the data cube. All values are given
  in the rest-frame of \object, and each colour bar has the same
  velocity scale. Some of the structure in the width of the
  Lyman-$\alpha$ emission lines is due to the fact that the
  Lyman-$\alpha$ emission has two separate components (see section
  \ref{subsec:emlines}). The poor positioning of the object within the
  IFU field-of-view is due to an error in the execution of the
  observation request. The orientation in each of the figures is the
  same as in figure \ref{fig:lyman_cont}. }
\label{fig:maps}
\end{figure*}

To better constrain the emission line characteristics of \object, we
extracted several wavelength regions of particular interest from the
datacube, corresponding to the regions exhibiting redshifted
Ly$\alpha_{1216}$, C{\sc~iv}$_{1549}$, and He{\sc~ii}$_{1640}$
emission. The emission lines in each wavelength region of interest
were fit with a one-dimensional Gaussian profile, using a $\chi^2$
minimisation procedure which allowed the central wavelength, linewidth
and normalisation to vary. The spectra were weighted using a 2D
gaussian profile of 5$\times$5 pixels centred around each spaxel to
account for the atmospheric seeing ($\sim$0.6\asec $\approx$3 spaxels)
in these data, and smoothed by 3 pixels ($\sim 4.1$\AA) in the
wavelength direction. To determine whether an emission line was
present in a given spaxel, we calculated values of $\chi^2$ for both a
flat continuum, and a single Gaussian emission line profile. We
required a minimum $\delta \chi^2$ of 25 (S\slash N of 5) to detect an
emission line. In this way, all of the non-blank emission line spaxels
in figure \ref{fig:maps} are detected at $>$ 5$\sigma$. The results of
the fitting procedure (intensity, central wavelength offset and
velocity width) are shown in Fig.~\ref{fig:maps}.

It is apparent that the emission lines have their maximum intensity
away from the continuum peak. The C{\sc~iv} line is, in fact,
undetectable at the continuum peak due to the deep absorption trough.
We place limits of 360$\pm$210 km\slash s, on any possible velocity
shear of the Lyman-$\alpha$ emission line over $\sim$25 kpc in
projection. Ly$\alpha$ (where the signal-to-noise ratio is the
highest) shows only a small region, roughly a square arcsecond in
size, where the gas is redshifted by about 100\,km\,s$^{-1}$,
approximately 0.8$''$ to the North-East of the continuum peak; this
feature is not seen in He{\sc~ii}, despite both lines showing the same
increase in velocity dispersion close to the continuum peak.

\begin{figure*}
\centering
\includegraphics[width=1.95\columnwidth]{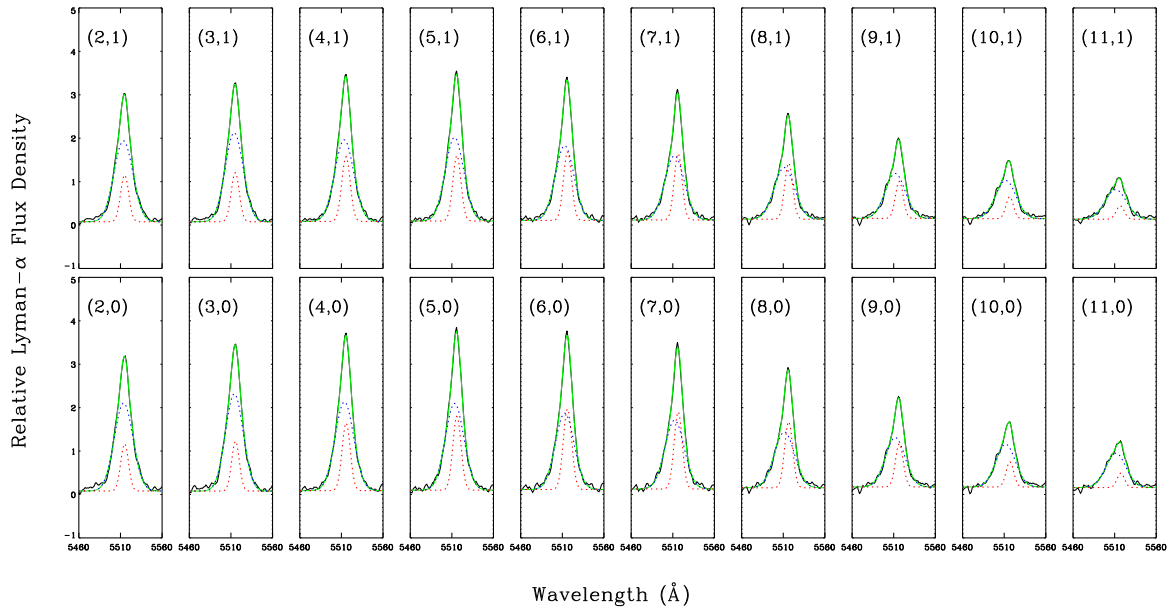}
\caption{The results of fitting our double-Gaussian model to the
  Lyman-$\alpha$ emission in \object. The data are shown as the black
  solid line with the spaxel indices displayed in the top left corner
  of each plot, whilst the best fit model (green dotted line) is a
  linear summation of the blue and red dotted components, whose
  central wavelengths and widths are shown in angstrom on the left and
  right sides of each plot, respectively. The two best-fit models vary
  remarkably little over the nuclear region of \object. Spaxel numbers
  count from (0,0) at the bottom left of the IFU as shown in
  Fig.~\ref{fig:maps}, with the peak of the continuum being at
  (10,0).}
\label{fig:lyacoreoverlay}
\end{figure*}

Unlike the profiles of the He{\sc ii} and C{\sc iv} emission lines,
which are consistent with a single gaussian component at $z = 3.534$,
it is clear that the Ly$\alpha$ emission line profiles measured in our
data cubes consist of two components. To investigate this further, we
fit the Ly$\alpha$ emission in \object\ with two Gaussian profiles and
the results are shown in Fig.~\ref{fig:lyacoreoverlay}. The model is a
good approximation to the observed profiles and, although the relative
strengths of the two components vary between spaxels, their redshifts
are virtually constant, at $z=3.5334 \pm 0.0008$ and $z=3.5363 \pm
0.0005$ for the two components. Their velocity dispersions are also
almost constant, with the Gaussians having FWHMs of 11.5~$\pm
~0.5$~\AA\ (625 $\pm$ 30 km\,s$^{-1}$) and 4.8~$\pm ~0.2$~\AA\ (260
$\pm$ 10 km\,s$^{-1}$), respectively. This decomposition also
trivially explains the structure seen in Fig.~\ref{fig:maps}. The
increased redshift to the NE of the continuum peak is due to an
increased contribution from the redder velocity component, while the
increased velocity dispersion is due to the bluer component (which
dominates the total flux, and is also broader) being broadest at this
location.


\subsection{The Absorption Lines}
\label{subsec:abslines}

Although the spectrum of \object\ displays very broad absorption
features similar to those of a broad absorption line quasar (BALQSO),
with the C{\sc~iv} trough extending blueward to
$\sim7000\rm\,km\,s^{-1}$, more species are seen in absorption in
\object\ than in BALQSOs. In addition to the C{\sc~iv} and Si{\sc~iv}
troughs, sharper absorption features are seen just shortward of the
predicted wavelengths of O{\sc~i}$_{1302}$ and
C{\sc~ii}$_{1335}$. Such features were also seen in 4C~41.17 (Dey et
al.\ 1997) and nearby star-forming galaxies such as NGC~1741 (Conti,
Leitherer \& Vacca 1996).

To better understand the relationship between the star-forming
region(s) and the AGN, we attempted to derive a redshift for the
starburst by Fourier cross-correlating a spectrum extracted from the
core of \object\ with that of NGC~1741 (Conti et al.\ 1996). Fourier
cross-correlation exploits the easily-derived fact that the Fourier
transform of the sum of two functions (here, spectra) is equal to the
sum of the transforms of the two functions. For this analysis, only
the region between the Ly$\alpha$ and He{\sc~ii} lines was used, to
avoid problems with the strong emission lines (C{\sc~iv} is not
observed at the continuum peak of the source; Fig.~\ref{fig:maps}).
The {\sc iraf} task {\tt fxcor} was used to perform the
cross-correlation, returning the amplitude of the correlation function
at a range of discrete velocities.

\begin{figure*}
\centering
\includegraphics[width=1.95\columnwidth]{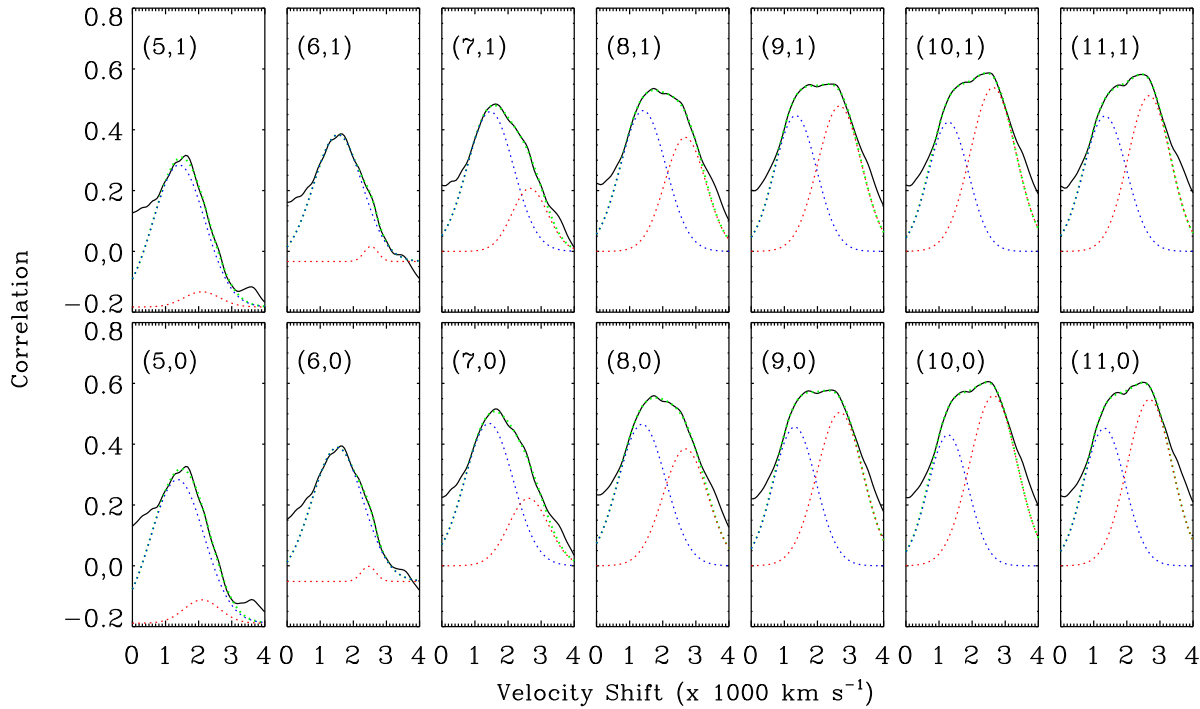}
\caption{The results of our Fourier cross-correlation between the core
  spaxels of \object\ (with the positions shown by the pixel
  co-ordinates in the top left of each frame), and the template
  spectrum of typical starburst galaxy NGC1741 (redshifted to $z=3.5$
  before cross-correlating). The peaks in the correlation function
  (black solid line) have been approximated by a best-fit model (green
  dotted lines), which is the sum of a a flat continuum and two
  Gaussian components (blue and red dotted lines).}
\label{fig:fxcor_result_fitting}
\end{figure*}

The cross-correlation function suggests the presence of two absorption
systems, separated by 1300 $\pm$ 200 km~s$^{-1}$ along the line of
sight. The results of this modelling are shown in
Fig.~\ref{fig:fxcor_result_fitting}. The velocities of the two
components, corresponding to redshifts of 3.5203 $\pm$ 0.0011 and 3.5401
$\pm$ 0.0027 (where the errors are derived based on the standard
deviation of the individual spaxel redshifts), suggesting that the
continuum emission observed in \object\ is the result of two
interacting starburst galaxies, moving relative to one another with a
line-of-sight velocity of $\sim 1300\rm\,km\,s^{-1}$.

\begin{figure}
\centering
\includegraphics[width=0.95\columnwidth]{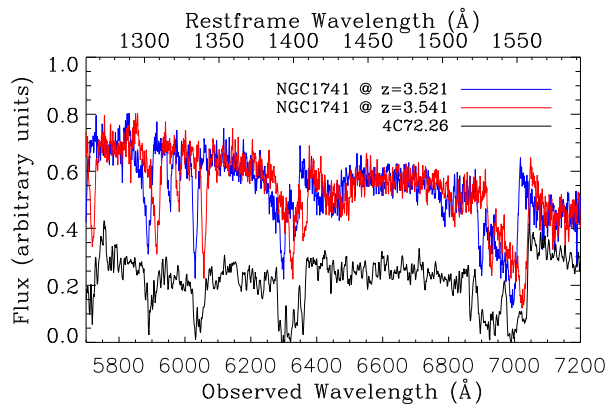}
\caption{A comparison between the observed spectrum at the core of
  \object\ and the spectrum of NGC~1741 at two separate redshifts
  corresponding to the best-fit redshifts from our Fourier
  cross-correlation (blue and red, corresponding to the lower- and
  higher-redshift systems, respectively). Note how two absorption
  systems are able to reproduce the structure in the C{\sc~ii} and
  C{\sc~iv} troughs shown in the spectrum of \object\ (shown in
  black).}
\label{fig:comparison}
\end{figure}

To confirm this interpretation, we have constructed a spectrum of two
star-forming regions using NGC~1741 as a template. We fit a smooth
continuum to the unabsorbed regions of the NGC~1741 and derive the
optical depth as a function of wavelength, $\tau(\lambda)$. A spectrum
can then be constructed as
\[
I(\lambda) = e^{-\tau(\lambda/(1+z_1))} (1 + f
e^{-\tau(\lambda/(1+z_2))} )
\]
where $z_1$ and $z_2$ are the redshifts of the star-forming regions,
and $f$ is the relative luminosity of the background system, compared
to the foreground one. This assumes that the foreground system
completely covers the background one, which appears to be the case
since the observed flux drops to zero in the C{\sc~iv} absorption
trough. To achieve this, we had to increase the optical depth in the
two systems, compared to that derived from the NGC~1741 spectrum. And,
as shown in Fig.~\ref{fig:comparison}, this deep trough is seen at a
wavelength consistent with the lower-redshift ($z=3.520$) system,
indicating that it must lie in front of the higher-redshift
($z=3.540$) one.
The depth of the C{\sc~iv} trough produced by the background (redder)
system sets $f\approx5$, since the continuum here is due to the
foreground system only.

We can also use the rest-frame ultra-violet continuum luminosity to
estimate the star formation rate associated with \object\ using the
calibration of Madau, Pozzetti \& Dickinson (1998). Since we have
clearly missed some of the continuum in our IFU observations, we have
rescaled the spectrum to make its flux scale consistent with the
\textit{HST\/} F702W measurement of 8.18\,$\mu$Jy (in a 4-arcsecond
aperture). A factor of 1.75 upward correction is required, which is
reasonable since the compact core lies very close to the edge of the
IFU's field of view (Fig.~\ref{fig:fov}) and therefore seeing will
redistribute almost half the flux outside our datacube. We derive a
luminosity density at 1500\,\AA\ of $3.1 \pm 0.5
\times10^{35}$\,W\,\AA$^{-1}$ and hence a star-formation rate of
300\,\Msolar\,yr$^{-1}$ (Salpeter IMF) or 680\,\Msolar\,yr$^{-1}$
(Scalo IMF), with a $\sim$16\% uncertainty associated with each
value. Although these numbers differ by a factor of two, they are both
an order of magnitude lower than that derived from the far-infrared
luminosity ($4000\pm700$\,\Msolar\,yr$^{-1}$) or PAH emission
($\sim8000$\,\Msolar\,yr$^{-1}$; Seymour et al.\ 2008).


We can also derive the SFR based on the starburst models of Leitherer
et al.\ (1995), in a manner similar to that demonstrated in Dey et
al.\ (1997) for 4C~41.17. As with 4C~41.17, we find evidence for
S{\sc~v}$_{1502}$ absorption, although the presence of two superposed
absorption systems results in a large width of the feature ($\sim
80\pm40$\AA). This, coupled with its low equivalent width, may explain
why it is not apparent in the Keck spectrum of De Breuck et
al. (2001), which was extracted over a larger spatial region. As this
feature is only seen in O stars (Walborn, Nichols-Bohlin \& Panek
1985; Walborn, Parker \& Nichols 1995), it implies a young stellar
population. However, a single power-law fit to the continuum of the
top panel of Fig.~\ref{fig:spectrum_both} over the range
5800--8000\,\AA\ gives $F_\lambda \propto \lambda^{-0.51\pm0.05}$,
implying either substantial reddening or an additional red spectral
component; we return to this point in Section~\ref{sect:sed}. The
Leitherer et al.\ models suggest that a continuously star-forming
population with a rate of 1\Msolar\,yr$^{-1}$ has a luminosity at
1500\,\AA\ of between $\log (L_{\rm 1500} /{\rm W\,A^{-1}}) = 32.25 -
33.16$ for ages of 1--9\,Myr. which corresponds to a star formation in
the range 210--1700\Msolar\,yr$^{-1}$. Despite a broad range of
estimates of the star-formation rate in \object, it is clear that a
massive starburst is underway with a rate of
$\sim$1000\Msolar\,yr$^{-1}$, comparable with the most extreme rates
seen in the Universe (e.g. in SCUBA galaxies, Smail, Ivison \&\ Blain,
1997, or Ivison et al., 2007). In both cases, approximately 80\% of
the total star formation occurs in the background galaxy.


\subsection{K-band spectroscopy of \object}

\begin{figure}
\centering
\includegraphics[width=0.90\columnwidth]{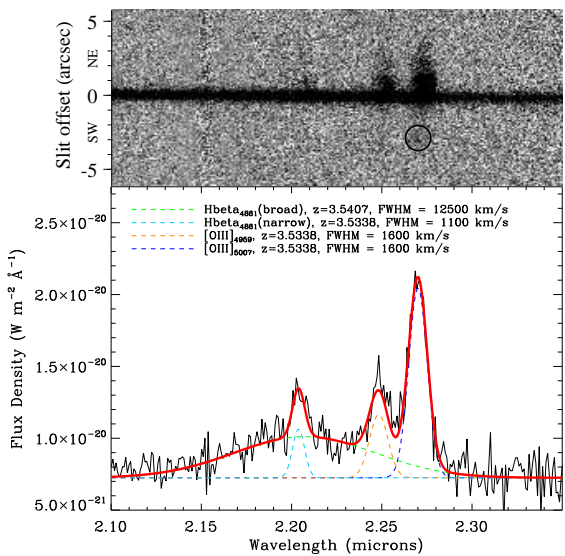}
\caption{CISCO $K$-band spectroscopy of \object, showing
  spatially-extended narrow emission lines of H$\beta$, and
  [O{\sc~iii}]. Top: two-dimensional spectrum, with [O{\sc~iii}]
  emission visible on both sides of the unresolved continuum. The
  $K$-band continuum peak is coincident with the optical peak, and
  therefore the assumed position of the radio core. The circle
  indicates the location of the [O{\sc~iii}] line emission discussed
  in the text. Bottom: one-dimensional spectrum, extracted through a
  1-arcsecond aperture, with five-component fit overlaid. The five
  components are flat continuum level of
  $7.2\times10^{-21}$W~m$^{-2}$~\AA$^{-1}$, broad H$\beta$ (green,
  dashed line), narrow H$\beta$ (light blue dotted-dashed),
  [OIII]$_{4959}$ (orange dot-dashed), and [OIII]$_{5007}$ (dark blue
  dotted). }
\label{fig:kspec}
\end{figure}

Fig.~\ref{fig:kspec} shows our CISCO $K$-band spectroscopy of
\object. In order to study the emission lines in \object, we
approximated the spectrum using a simple model consisting of four
Gaussian components and a continuum component that is flat in
$F_\lambda$. In fitting the spectrum with this simple model, we
required the redshifts of the narrow \Hbeta\ and [O{\sc~iii}]
components to be equal since they arise from the same emission-line
gas, whilst the broad component's central wavelength was allowed to
vary. The widths of the two [O{\sc~iii}] lines were set to be equal,
and their flux ratio constrained to be 3. The best fit model is shown
in Fig.~\ref{fig:kspec}; we find that our simple model adequately
describes our data; the broad component of the \Hbeta\ emission has a
best-fit redshift of $z=3.5407~\pm~0.0066$, while the narrow emission
lines are blueshifted by $\sim 460\rm\,km\,s^{-1}$, although this is
approximately equal to the $1\sigma$ error on the redshift of the
broad H$\beta$ component, so the two redshifts are formally
consistent.

With the results of our model fitting, we can use the calibration from
McLure \& Jarvis (2002) to determine the mass of the central black
hole from the FWHM of the H$\beta$ emission line:

\begin{equation}
\frac{M_{\rm bh}}{\Msolar} = 4.74 \left( \frac{\lambda L_{5100}}{10^{37}{\rm W}}\right)^{0.61} \left[ \frac{{\rm FWHM(H}\beta{\rm)}}{{\rm km~s}^{-1}} \right]^2.
\label{mj02}
\end{equation}

\noindent The derived value of $(1.8\pm0.2) \times
10^{10}$\Msolar\ (the error quoted is from the fit) is larger than the
most massive black holes in the McLure \& Jarvis (2004) sample,
although the 1$\sigma$ uncertainty in the calibration is a factor of
2.7. However given that \object\ was originally discovered through
low-frequency radio emission we might expect that it is aligned closer
to the plane of the sky than the majority of optically selected
quasars. If the broad-line region is akin to a rotating disk then we
would expect, on average, broader emission lines (see e.g. Jarvis \&
McLure, 2006). Therefore, although our black-hole mass estimate is
subject to uncertainties of order a factor of 5, we can say with
certainty that we are observing an extremely massive black hole.

We note that the line-emitting gas is only present on the North-East
side of \object. We associate this with the foreground since there is
a tentative one-sided jet at radio wavelengths and this is the longer
radio arm, suggesting a shorter light travel-time to the observer,
although doubt remains as to whether this is in fact the
case. McCarthy, van Breugel \& Kapahi (1991) suggested that
emission-line asymmetries were due primarily to differences in the
density on the two sides of the nucleus, supporting this with evidence
of a correlation between the side with the strongest line emission and
the side with the shortest radio arm. However, this is not the case
for \object\ where the approaching arm is the longer of the two, and
so we instead suggest that there is an extensive dust disc obscuring
the far side of the emission-line region. The near-side line emission
extends for $\sim2.5''$ (18\,kpc) from the nucleus, and close
inspection of the longslit spectrum reveals faint line emission
located $\sim2.5"$ from the nucleus on the far side of the source,
which we speculate is material just beyond the extent of this putative
disc (figure \ref{fig:kspec}). We also examined an archival WHT-ISIS
spectrum of \object, which shows the Lyman-$\alpha$ emission extending
for $\sim$8\asec\ in the NE direction, without any emission being
detected on the SW side of the core; this is not surprising given the
ease with which Lyman-$\alpha$ photons may be absorbed. Comparing the
surface brightness of the [O{\sc~iii}] emission on both sides of the
nucleus, we conservatively estimate an asymmetry of $>7$ (3$\sigma$),
implying $A_V>1.9$ (adopting the SMC extinction law of Pei
1992). Assuming the relationship between gas column density and
extinction derived by Bohlin, Savage \& Drake (1978), we calculate
$N_{\rm H}>3.5 \times 10^{21}\rm\,cm^{-2}$. If this column persists
across an entire 18-kpc-radius disc, then the total mass of gas
contained within it is $M_{\rm gas}>3\times10^{10}\Msolar$, which
would be sufficient to power the starburst at its present rate for at
least a few million years. This estimate is consistent with the value
of $4.5\times10^{10}\Msolar$ of H$_2$ conservatively estimated by
Papadopoulos et al.\ (2000) from their CO(4--3) observations. Our
proposed extent for the disc is marginally consistent with the upper
limit of 4$''$ for the extent of the CO, for which Papadopoulos et
al.\ derive a dynamical mass of $6\times10^{11} (\sin i)^{-2}\Msolar$.

We can also use the ratio of the fluxes in the broad and narrow
components of the H$\beta$ line to estimate the attentuation to the
broad line region within \object. If we assume an intrinsic ratio of
broad to narrow H$\beta$ of 40 (e.g.\ Jackson \& Eracleous 1995), then
the ratio determined from our four-component Gaussian fit corresponds
to a factor of 4.1 in extinction at 4861\,\AA, equivalent to $A_V
\approx 1.4$, although this is highly uncertain given the available
data.

\section{Spectral Energy Distribution of \object}
\label{sect:sed}

\begin{table*}
\centering
\caption{Photometry of \object\ used in our SED fitting analysis
  (Fig.~\ref{fig:sed}). These data are compiled from several different
  publications as shown in the table. The errors are quoted to
  1$\sigma$; where limits are given, they correspond to 3$\sigma$. }
\vspace{0.15cm}
\begin{tabular}{|l|l|l|l|}
\hline
Photometric Band &  Flux Density ($\mu$Jy) & Instrument & Reference\\
\hline
\hline
F702W & 8.18 $\pm$ 0.16 & WFPC2 & This work \\
$K_{\rm s}$ & 134 $\pm$ 7 & CISCO & This work \\
$K_{\rm s}$ (line-corrected) & 109 $\pm$ 11 & CISCO & This work \\
3.6$\mu$m & 200 $\pm$ 20 & IRAC & \multirow{4}{*}{Seymour et al., 2007} \\
4.5$\mu$m & 229 $\pm$ 23 & IRAC &  \\
5.8$\mu$m & 241 $\pm$ 25 & IRAC &  \\
8.0$\mu$m & 480 $\pm$ 48 & IRAC &  \\
12$\mu$m & 840 $\pm$ 100 & ISOCAM & Siebenmorgen et al., 2004 \\
16$\mu$m  & 1320 $\pm$ 70 & IRS & Seymour et al., 2008 \\ 
24$\mu$m  & 1910 $\pm$ 100 & MIPS & \multirow{3}{*}{Seymour et al., 2007} \\ 
70$\mu$m  & 16200 $\pm$ 1900 & MIPS &  \\ 
160$\mu$m  &  $<$63300 & MIPS &  \\ 
350$\mu$m  &  90000 $\pm$ 15000  & SHARC-{\sc ii} & Greve et al., 2006 \\ 
450$\mu$m  &  33000 $\pm$ 17000 & SCUBA & Reuland et al., 2004 \\ 
850$\mu$m  &  13500 $\pm$ 1300 & SCUBA & \multirow{2}{*}{Papadopoulos et al., 2000} \\ 
1250$\mu$m  & $<$3000 & PdBI &  \\
\hline
\end{tabular}
\label{photometry}
\end{table*}

\begin{figure*}
\centering
\includegraphics[width=1.83\columnwidth]{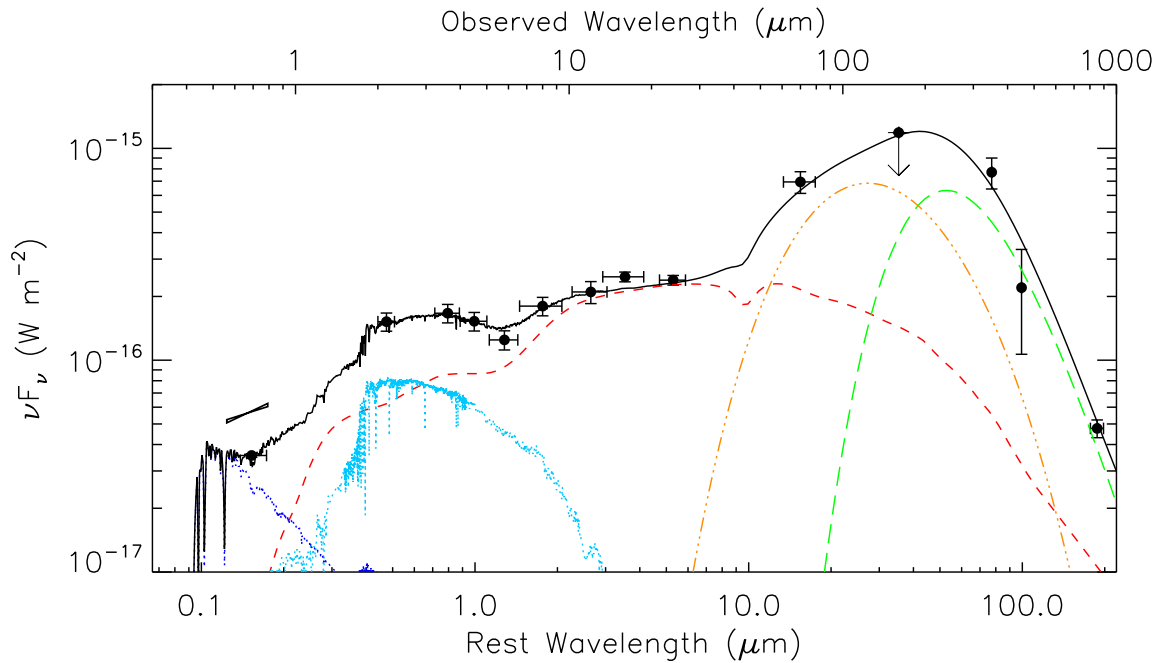}
\caption{Spectral energy distribution of \object; the best fit model
  (solid line) is the sum of five components: a 30-Myr old starburst
  population (blue dotted line), a 1-Gyr stellar population (light
  blue dotted line), a radio-loud QSO template (red dashed), a warm
  dust component (orange dot-dashed) and a modified grey-body cold
  dust profile (green dashed) taken from Seymour et al. (2008). The
  bow-tie above the spectrum in the rest-frame ultraviolet indicates
  the slope of the continuum as measured from the spectrum. For
  details of the fitting and the photometry see section \ref{sect:sed}
  and table \ref{photometry}.}
\label{fig:sed}
\end{figure*}

We have compiled archive photometry of \object, similar to the results
of Seymour et al.\ (2008), in Table~\ref{photometry}. However our new
observations demonstrate that the continuum is dominated by a QSO at
near-infrared wavelengths, and a luminous starburst in the optical/UV,
which were not included in the Seymour et al. SED analysis.

The luminosity of the QSO component is well constrained by the
10--20-$\mu$m photometry (observed frame), since this wavelength range
is redward of the peak of stellar emission, and blueward of the
thermal emission from the starburst-heated dust. Demanding that the
model QSO component (a median radio-loud QSO model from Elvis et al.,
1994) alone fits the photometry in this region also provides strong
constraints on the amount by which it can be extinguished. The F702W
photometry requires $A_V>0.7$ or else the QSO emission alone will
exceed the total flux.  The rest-frame equivalent width of the broad
H$\beta$ line is 86\,\AA, and for this to be consistent (at the
2$\sigma$ level) with the distribution of local, bright QSOs from
Miller et al.\ (1992) requires that at least 41\,per cent of the
$K$-band continuum light is from the QSO, and this in turn demands
$A_V<1.2$. This relatively narrow window of acceptable extinction
values is even smaller if an LMC or Milky Way extinction law is used,
since they are shallower in the ultraviolet and hence a larger visual
extinction is needed to produce the required UV attenuation.

With these constraints, the reddened QSO still contributes to the
rest-frame ultraviolet emission, and explains why the spectrum is much
redder than would be expected for a young stellar population
(Section~\ref{subsec:abslines}). Dey et al. (1999) noted that the
ultraviolet continuum emission in \object\ was unpolarised, consistent
with our SED analysis in excluding a dominating scattered QSO
component at these wavelengths. In fact, the young stellar population
(which we model with a 30-Myr-old simple stellar population from the
BC03 models of Bruzual \& Charlot 2003) must be unreddened for the
overall spectral shape to remain consistent with our GMOS
observations. The luminosity of the young stellar component is
therefore tightly constrained by the F702W photometry.

As the QSO must contribute between 40--70\,per cent of the $K$-band
continuum light, and the young stellar population only contributes
$\sim10$\,per cent, there must be a further component which provides
20--50\,per cent of the continuum light. Since this component must
peak at $\lambda_{\rm rest}\sim1\,\mu$m, we assume it is stellar in
origin, and a natural interpretation is a more evolved population of
stars. 

However, we note that this component alone must have $17.7<K<18.7$,
which places it between 1.1 and 2.1 magnitudes brighter than the
polynomial fit to the $K$--$z$ relation of Willott et
al.\ (2003). However, we note that this is consistent with the
distribution of $K-$band magnitudes at $z>1.8$ in Jarvis et al. (2001,
and more recently those in Bryant et al., 2009) where there appears to
be a slight turnover in the $K$--$z$ relation. Furthermore, the
typical scatter of 0.6 magnitudes across the whole of the $K$--$z$
diagram and the correlation between host galaxy mass and radio
luminosity previously discussed means that it is not unreasonable to
observe such a magnitude for this particularly powerful high-redshift
radio source. As the $K$--$z$ relation closely follows the passive
evolution of a present-day 3$L^*$ galaxy that formed at a redshift
$z_{\rm f}=10$, this stellar component would evolve into a very
luminous (8--20$L^*$) galaxy by the present epoch, unless it formed
more recently. We, somewhat arbitrarily, use an unreddened 1-Gyr-old
population, whose age corresponds to $z_{\rm f}=6.7$. The mass of
these stars required to fit the SED is $1.8\times10^{12}\Msolar$
(assuming a Initial Mass Function from Chabrier, 2003, and Solar
metallicity), although this is not very sensitive to the assumed age,
since younger populations, although they possess an intrinsically
smaller mass-to-light ratio, must be reddened in order to maintain the
overall shape of the SED. Unfortunately, we lack the detailed spectral
coverage in the 1--2\,$\mu$m range to remove the degeneracy between
age and extinction for this population, and we can obtain acceptable
fits with a range of values, although all require a mass of
$>1\times10^{12}\Msolar$. Populations with exponentially-declining
star formation rates can also fit this population, provided the
$e$-folding time is less than the age and they are reddened; an
additional unreddened young population is also required, however.

In summary, our best-fit spectral energy distribution consists of five
components:
\begin{itemize}
\item a 30-Myr-old simple stellar population from Bruzual \& Charlot
  (2003), with a mass of $2\times10^{10}\Msolar$,
\item a 1-Gyr simple stellar population, also from Bruzual \& Charlot
  (2003), with a mass of $1.8\times10^{12}\Msolar$,
\item a median radio-loud QSO template from Elvis et al.\ (1994) which
  suffers 1.2 magnitudes of extinction, according to the Small Magellanic
  Cloud law of Pei et al.\ (1992)
\item a warm dust component described by a power-law
  with an index of two and exponential cutoffs at high and low
  frequency,
\item a modified black body with a temperature of 50K and emissivity
  index $\beta=1.5$ (following the results of Seymour et al.\ 2008).
\end{itemize}

It is clear from Fig.~\ref{fig:sed} that the overall SED is readily
split into the ultraviolet/optical/near-infrared regime (fit by the
first three components listed above), and the far-infrared region (fit
by the last two components), with effectively no cross-talk between
these two regimes. Our fit to the far-infrared region is very similar
to that of Seymour et al.\ (2008), while our fit over the two decades
in wavelength from 0.1--10\,$\mu$m is constrained by nine photometric
measurements, plus information from our spectroscopy.

We note that our fit significantly underestimates the 16-$\mu$m flux
measured by \textit{Spitzer\/}, but the IRS spectrum of
\object\ clearly shows strong PAH emission which is not included in
the QSO SED (Seymour et al.\ 2008). The need to include warm dust
emission in addition to the QSO SED (whose far-infrared continuum is
also produced by thermal dust emission) indicates a substantial extra
heating source, consistent with the massive starburst observed in our
GMOS spectrum. Finally, we note the good agreement between our
estimates of the foreground extinction to the QSO derived from the
H$\beta$ emission ($A_V=1.4$) and the SED ($A_V=1.2$).

\section{Discussion}

\begin{figure}
\centering
\includegraphics[width=0.95\columnwidth]{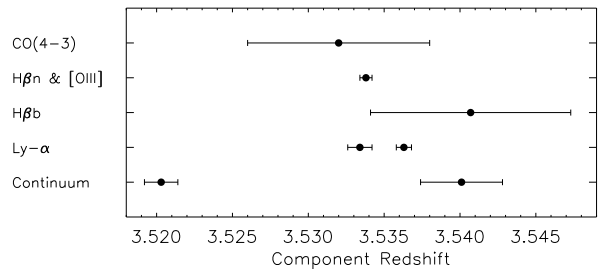}
\caption{A comparison between the redshifts derived for each of the
  different observables. The Lyman-$\alpha$ redshifts and those of the
  Fourier cross-correlation analysis between the continuum region of
  \object\ and the NGC1741 template are shown on respective rows, and
  the locations of the emission lines from our Subaru spectrum of
  Fig.~\ref{fig:kspec} are also shown, compared with the CO$(4-3)$
  redshift from Papadopoulos et al., 2000. The error bars correspond
  to the 1$\sigma$ uncertainties based on our data. }
\label{fig:redshiftcomparison}
\end{figure}

A comparison of the results of our Ly$\alpha$ fitting and our Fourier
cross-correlation of the continuum properties of \object\ are shown in
Fig.~\ref{fig:redshiftcomparison}. The redshift of the broad H$\beta$
component is consistent with the redshift of the redder (background)
starburst, at $z=3.5407 \pm 0.0066$, and we identify this as the
redshift of the radio-loud AGN. Although both Ly$\alpha$ emission-line
systems are formally consistent with this redshift (due to the large
uncertainties associated with the continuum and broad-line redshifts),
we identify them as the approaching sides of two bidirectional
outflows, with the receding sides hidden by the large-scale dust disc
described in Section 3.3. This asymmetric obscuration can also explain
the lack of spatial coincidence between the peaks of the Ly$\alpha$
and continuum emission.

The more blueshifted of the Lyman- components ($z = 3.5334$) has a
redshift consistent with that measured from the narrow H$\alpha$ and
[O{\sc~iii}] lines, and also the C{\sc~iv} and He{\sc~ii} emission,
and the richness of this emission-line spectrum indicates that it is
associated with the AGN activity. The less strongly blueshifted (and
narrower) Ly$\alpha$ emission-line system ($z=3.5363$) is not
associated with any other emission lines and we infer that this is an
outflow being driven at a line-of-sight velocity of
$\sim250$\,km\,s$^{-1}$ by the starburst in the radio galaxy host.
The lower-redshift (foreground) starbust ($z=3.5203$) has no
detectable line emission associated with it, although we note that
around 25\%\ of starbursting galaxies are not associated with an
extended Ly$\alpha$ halo (Pettini et al.\ 1997), so this is not
especially surprising.

It is clear that the supermassive black hole and most vigorous
star-forming region both lie at $z=3.540 \pm 0.003$, while the CO(4-3)
emission has a redshift $z=3.532 \pm 0.006$ (Papodopoulos et
al.\ 2000), which is formally consistent with the higher redshift
starburst. However, it is reasonably common for the molecular gas to
end up between two galaxies undergoing collisons (e.g. Zhu et al.,
2007, and also Ivison et al., 2008), which may explain the slight
discrepency.

The commonly-proposed scenario of radio jet-induced star formation,
which has been proposed for rapidly star-forming radio galaxies, does
not appear to be appropriate for \object. We note the lack of
similarity between this source and 4C~41.17, which has been proposed
as a jet-induced starburst (Dey et al.\ 1997; Bicknell et al.\ 2000);
that source displays extended ultraviolet emission with morphological
similarities to the radio structure, whereas \object\ has a very
compact UV continuum; and the velocity field of 4C~41.17 is disturbed
whereas that of \object\ is smooth. In addition, the analysis in
Section~\ref{subsec:abslines} has clearly demonstrated the presence of
two star-forming systems aligned closely to the line of sight, yet the
radio jet is not pointing along this axis; the geometry therefore does
not favour the foreground system being influenced by the radio jet and
there is no need to demand that the background system is affected in
this way.

Nevertheless, we consider the jet-induced starburst scenario in the
manner of Bicknell et al.\ (2000). We first constrain the velocity of
the shock from the FWHM of the emission lines in the high-excitation
($z=3.534$) spectrum (400--500\,km\,s$^{-1}$) and their blueshift with
respect to the assumed systemic velocity (460\,km\,s$^{-1}$). This
argues for a much lower velocity shock than in 4C~41.17, and is
supported by the lack of C{\sc~iv} emission at the nucleus, since such
shocks produce very little emission from the precursor (Allen et
al.\ 2008), and almost all the C{\sc~iv} flux comes from the
post-shock region where it can be absorbed by the starburst-driven
outflow. A shock velocity of $\sim300$\,km\,s$^{-1}$ produces roughly
equal fluxes of C{\sc~iv} and He{\sc~ii}, as observed in the halo, and
C{\sc~iv}/C{\sc~iii}]$\sim$3.5, consistent with the weakness of the
  C{\sc~iii}]$_{1909}$ line (although this lies in a region strongly
    affected by night sky lines and therefore we cannot measure a
    reliable flux for it). With the caveat about the night sky lines,
    we can assume with confidence that C{\sc~iv}/C{\sc~iii}]$>$2.0,
      and compare these values with the results of Bryant et al. (2009
      -- their figure 13). These emission line ratios may readily be
      explained by photoionisation according to their simple model,
      and this conclusion is not altered by assuming even considerably
      weaker C{\sc~iii}] emission than our deliberately conservative
        upper limit.

The absence of vigorous star formation in the halo region could be
explained by a much lower gas density, since the time-scale for
gravitational instability increases with decreasing density
(equation~9 of Bicknell et al.\ 2000; after Elmegreen \& Elmegreen
1978), and there may have been insufficient time for them to collapse
since the passage of the radio jet.

Using the results of Allen et al.\ (2008) and the scaling of Dopita \&
Sutherland (1996), we infer that the total H$\beta$ luminosity of a
shock with velocity $300v_{300}\rm\,km\,s^{-1}$ propagating through a
medium of density $n\rm\,cm^{-3}$ is $L_{\rm H\beta} \approx
1.7\times10^{32} v_{300}^{2.41} n\rm\,W\,kpc^{-2}$. For the specific
case of a 300\,km\,s$^{-1}$ shock, this produces a He{\sc~ii}
luminosity $L_{\rm HeII} = 2.0 \times 10^{32} n\rm\,W\,kpc^{-2}$.

Integrating the He{\sc~ii} flux over a $1.6\times1.6$\,arcsec$^2$
($11.3\times11.3$\,kpc$^2$) region of the halo produces a value of
$f_{\rm HeII} = 8.0\times10^{-20}\rm\,W\,m^{-2}$. We note that we are
likely to have missed some flux but do not apply any correction to
account for this. The luminosity of this line is $1\times10^{36}$\,W,
which requires $nA_{\rm sh}\approx5000$, where $A_{\rm sh}$ is the
area of the shock in kpc$^2$. Since $n\la1$ is needed to prevent
star-formation on short time-scales, this result is inconsistent with
the area over which we have measured the line flux.

Similarly, we can consider the situation in \object's compact core,
where the bulk of the star formation occurs. Here, the observed
He{\sc~ii} luminosity is $5\times10^{35}$\,W, although it is certain
that significant emission has been lost due to the poor positioning of
the IFU. Since the core is $<500$\,pc in size, we require a pre-shock
density $n>10^4$\,cm$^{-3}$. This would result in a recombination
density $>10^6$\,cm$^{-3}$ and hence significant suppression of the
[O{\sc~iii}] doublet through collisional de-excitation, for which we
see no evidence. In truth, the core density would need to be much
higher, since the unresolved core must also include the transition
region where the density drops from the high core value to the much
lower value in the halo which prevents the collapse of clouds (and
hence star formation) in the halo region, and the density is unlikely
to vary discontinuously. Further analysis would require assumptions
about a density profile in the neighbourhood of the core, which is
beyond the scope of this paper.

Considering these facts, we therefore conclude that jet-induced star
formation does not provide a realistic explanation for \object.
Instead, the kinematics of the two galaxies suggests that there has
been a collision, such as that proposed for the $z\sim1$ source 3C~356
(Simpson \& Rawlings 2002), that has triggered the AGN activity and
the star formation.

We finally note that, despite the extreme starburst we are witnessing
in \object, the rest-frame UV/optical/near-IR spectral energy
distribution requires a massive stellar population with an age of a
few hundred Myr to already be in place. The massive starbursts found
in high-redshift radio galaxies might therefore not be directly
related to the major assembly of their stellar mass, but instead due
to the presence of a large gas reservoir, absent in lower-redshift
systems, within which star formation is induced by a mutual trigger,
although jet-induced effects may be important in individual cases such
as 4C~41.17 (Bicknell et al., 2000).

\section{Conclusions}
\label{conclusions}

We have analysed new spectroscopy of the $z\approx3.5$ `broad
absorption-line radio galaxy' \object, and found it to be associated
with a system of two vigorously star-forming galaxies, separated from
one another by 1300 $\pm$ 200 km~s$^{-1}$, and the more active of
which hosts the supermassive black hole responsible for the extended
radio source. Both galaxies display P~Cygni-like absorption line
profiles, while the AGN host also shows Ly$\alpha$ emission indicative
of a galaxy-wide `superwind' (e.g., Steidel et al.\ 2000; Taniguchi \&
Shioya 2001; Matsuda et al.\ 2004; Smith \& Jarvis 2007, Smith et al.,
2008). The AGN host is also the source of a luminous, highly-ionized
outflow.

We have found tentative evidence for a 40-kpc dust disc, whose size
and mass are consistent with being the site of the CO emission
observed by Papadopoulos et al.\ (2000). From several lines of
argument, we have ruled out a jet-induced star-formation scenario for
\object, and instead proposed that the star formation is caused by a
collision between the two galaxies. The need for an old stellar
population suggests that even at $z=3.5$ we are not witnessing the
initial episode of star formation in massive elliptical galaxies.

\FloatBarrier

\section*{Acknowledgments}
The authors would like to thank Loretta Dunne for valuable
discussions. AMS acknowledges a Royal Astronomical Society Sir Norman
Lockyer Fellowship. This publication was based, in part, on
observations obtained at the Gemini Observatory, which is operated by
the Association of Universities for Research in Astronomy, Inc., under
a cooperative agreement with the NSF on behalf of the Gemini
partnership: the National Science Foundation (United States), the
Science and Technology Facilities Council (United Kingdom), the
National Research Council (Canada), CONICYT (Chile), the Australian
Research Council (Australia), Ministrio da Cincia e Tecnologia
(Brazil) and Ministerio de Ciencia, Tecnologa e Innovacin Productiva
(Argentina). It is also based on data collected at the Subaru
Telescope, which is operated by the National Astronomical Observatory
of Japan.

\bsp

\label{lastpage}

\end{document}